\documentclass[twocolumn,floatfix,preprintnumbers,aps,prl,showpacs,footinbib,amsmath,amssymb,amsfonts]{revtex4}

\usepackage{graphicx}

\usepackage[dvips,usenames]{color}
\usepackage{dsfont}
\usepackage{hyperref}

\newcommand{\MyDelta}{{\Delta}}

\newcommand{\Bary}{{\rm Bar}}

\newcommand{\be}{\begin{equation}}
\newcommand{\ee}{\end{equation}}
\newcommand{\bea}{\begin{eqnarray}}
\newcommand{\eea}{\end{eqnarray}}

\def\st{\sigma_{\mathrm T}}

\def\dd{{\rm d}}
\def\HH{\mathcal{H}}

\definecolor{Myblue}{rgb}{0,0,1}
\definecolor{Myred}{rgb}{1,0,0}

\def\iB{{i}}
\def\jB{{j}}
\def\kB{{k}}

\def\iT{{i}}
\def\jT{{j}}
\def\kT{{k}}

\def\ipl{{\rm pl}}
\def\ib{{\mathrm b}}
\def\ir{{\gamma}}
\def\ie{{\mathrm e}}

\def\is{{\mathrm s}}

\begin{document}
\title{The tight-coupling approximation for baryon acoustic oscillations}

\author{Cyril Pitrou}
\affiliation{
Institute of Cosmology \& Gravitation, University of Portsmouth, Portsmouth PO1 3FX, United Kingdom}

\date{\today}

\begin{abstract}
The tight-coupling approximation (TCA) used to describe the early dynamics of the baryons-photons system is systematically built to
higher orders in the inverse of the interaction rate. This expansion can be either used to grasp the physical effects by deriving simple analytic
solutions or to obtain a form of the system which is stable numerically at early times. In linear cosmological perturbations, we estimate numerically its precision, and we discuss the
implications for the baryons acoustic oscillations. The TCA can be extended to the second order cosmological perturbations, and in
particular we recover that vorticity is not generated at lowest order of this expansion. 
\end{abstract}
\maketitle


The baryon acoustic oscillations (BAO) generated when baryons and
photons were highly coupled have now been observed in the cosmic
microwave background~\cite{Bernardis:2000gy,Komatsu:2010fb} and in the large scale
structures~\cite{Eisenstein:2005su}. The shape and amplitude of these oscillations cannot
be obtained analytically and one resorts either to a numerical
resolution or to a less accurate WKB approximation on the full
system of dynamical equations for the cosmic fluids and the metric. On
the one hand the numerical integration has stability issues due to strong
restoring forces, and on the other hand the precision of the WKB approximation is limited. The
tight-coupling approximation, which is an expansion in the interaction
rate [see Eq.~(\ref{DefTCA}) below], is a way to avoid these two
issues~\cite{Peebles1970,Ma:1995ey}. The orders of the approximation
are denoted by TCA-$n$, and one can always consider the
equations up to TCA-$n$ with $n$ sufficiently large to ensure the required
accuracy, in the range of time for which the expansion is converging. In
this paper, we first recast the baryons-photons dynamics
into a total fluid system, and we then present the TCA expansion. We
also estimate numerically the convergence before recombination. A closed
form at TCA-1 for photons density perturbations is then derived from which the
main features of the dynamics are deduced. At second order in
cosmological perturbations, the TCA-0 is sufficient at early times
given the precision required for all practical purposes. 
An exception arises for vorticity since it is not generated for a
single perfect fluid, and we recover the known result that it is not
generated below TCA-1~\cite{Kobayashi:2007wd,Maeda2008}. Our analysis
is of interest for high precision BAO computations but also sheds some
light on the fluid approximation of the baryons-photons system in
general relativity and cosmology by offering another method for solving cosmological perturbations.

\subsection{Perturbations}
{\it Geometry}: We perturb the metric in the conformal Newtonian gauge according to
\be\label{metric}\dd s^2 =a^2\left[ (1-2 \Psi)\delta_{\iB\jB}\dd x^{\iB}\dd x^{\jB}-(1 + 2\Phi )\dd\eta^2 + 2
 S_{\iB} \dd x^{\iB}\dd\eta\right]\nonumber
\ee
where $a$ is the scale factor and $\eta$ the conformal time. $S_\iB$
is a vector type peturbation ($\partial_\iB S^\iB=0$) considered only
at second order in perturbations since vector
perturbations decay at first order if not sourced. We also do not consider the tensor perturbations.
A prime denotes a derivative with respect to $\eta$, and $\HH\equiv
a'/a$ .

{\it Baryons-photons system:} The stress energy tensor of a species labelled by $\is$
is decomposed as
\be
T_\is^{\mu\nu} = (\rho_\is+P_\is) u_\is^{\mu} u_\is^{\nu} + P_\is g^{\mu\nu}+P_\is\Pi_\is^{\mu\nu}\,.
\ee
The energy density is  perturbed according to \mbox{$\rho_\is=\bar
\rho_\is(1+\delta_\is)$}, velocities are perturbed according to
\mbox{$u_\is^{\iB}=1/a[(1+\Psi)v_\is^{\iB}-S^{\iB}]$} such that $v_\is^\iT$ matches the tetrad components of the velocity~\cite{Pitrou2008}.
We assume that baryons have no anisotropic stress ($\Pi^\ib_{\iB\jB}=0$), have equation of
state \mbox{$w_\ib\equiv \bar P_\ib/\bar \rho_\ib=0$}, whereas photons have equation of state $w_\ir=1/3$. The adiabatic speed of sound \mbox{$c_\is^2\equiv P_\is'/ \rho_\is'$} is constant for these two fluids and satisfies
\mbox{$c_{\ir/\ib}^2=w_{\ir/\ib} $}. In general for a perfect fluid, if $w_\is$ is not constant, $c_\is^2 \neq w_\is$ and \mbox{$w_\is'=-3\HH(c^2_\is-w_\is)(1+w_\is)$}.
For photons, free-streaming produces an anisotropic stress $\Pi^\ir_{\iB\jB}$ and in
general we have to consider the moments of the temperature
$\Theta_{\iB_1\dots \iB_\ell}$, and also the electric and magnetic
type multipoles ($E_{\iB_1\dots \iB_\ell}$ and $B_{\iB_1\dots \iB_\ell}$) to describe linear polarization. The lowest multipoles are related to the fluid quantities
up to second order by~\cite{Pitrou2008}
\be
\Theta_{\iB}=\frac{\rho_\ir v^\ir_\iB}{\bar \rho_\ir}\,,\quad
\Theta_{\iB\jB}=
\frac{5}{2}\left(\frac{\rho_\ir \Pi^\ir_{\iB\jB}}{4\bar \rho_\ir}+v^\ir_{\langle\iB} v^\ir_{\jB\rangle}\right)\,.
\ee
At first order, there are only scalar perturbations, the magnetic type
multipoles vanish and the temperature scalar multipoles are defined by \mbox{$\Theta_{\iB_1\dots \iB_\ell}
\equiv (-1)^\ell  (2 \ell-1)!/\ell!
\partial_{\langle\iB_1}\dots \partial_{\iB_\ell\rangle} \hat
\Theta_\ell$} (and similarly for electric multipoles) where $\langle
\dots \rangle$ means symmetric traceless part of the indices. In Fourier space, for a given mode ${\bf k}$ with
$k\equiv |{\bf k}|$, we define dimensionless multipoles from the previous ones by  $\Theta_{\ell}
\equiv k^{-\ell }\hat \Theta_{_\ell} $, and similarly for electric multipoles. For velocities, we use instead $v^\is_\iB
\equiv \partial_\iB \hat v^\is$ and $v_\is=\hat v_\is/k$, that is $v_\ir =
-\Theta_1$. 
\subsection{Baryons-photons system and tight-coupling}

{\it Fluid equations:} In Fourier space, the conservation and the
Euler equations at linear order are, with $\is=\ib,\ir$
\bea
\hspace{-0.1cm}{\rm C}_\is&\equiv&\left(\frac{\delta_\is}{1+w_\is}\right)'-k v_\is-3 \Psi'=0\\
\hspace{-0.1cm}{\rm E}_\is&\equiv&v_\is'+(1-3 c_\is^2)\HH
v_\is+k\Phi+\frac{kc_\is^2}{1+w_\is}\delta_\is-\frac{2k}{5}\Theta_2^\is={\cal C}_\is.\nonumber
\eea
The collision terms for baryons and photons are obtained from the
Boltzmann equation~\cite{Ma:1995ey} and are given by
\be\label{EqCollision1}
{\cal C}_\ir\equiv \tau'(v_\ib-v_\ir)\,,\quad {\cal C}_\ib\equiv -{\cal C}_\ir/R
\ee
with $R\equiv 3\bar \rho_\ib/(4 \bar \rho_\ir)$, and the interaction rate is \mbox{$\tau'\equiv a n_\ie \st$} where $n_\ie$ is the number density
of free electrons and $\st$ the Thomson cross-section. 

The plasma of photons and baryons is an unperfect fluid
(labelled by $\ipl$) whose energy density and velocity are given by
\be\label{DefrhoV}
\rho_\ipl\equiv \sum_{\is=\ib,\ir}\rho_\is\,,\quad(\rho_\ipl+P_\ipl)u_\ipl^\mu\equiv\sum_{\is=\ib,\ir}(\rho_\is+P_\is)u_\is^\mu.
\ee
We can infer easily the equation of state of the plasma, and by considering
its time evolution we find its adiabatic speed of sound. They read
\be
 w_\ipl=(3+4R)^{-1}\,,\quad c_\ipl^2=[3(1+R)]^{-1}\,\,.
\ee
If we define a reduced energy density contrast by
\mbox{$(1+w_\is){\MyDelta}_\is\equiv (1+w_\ipl) \delta_\is$}, we obtain from
Eqs.~(\ref{DefrhoV}) the very simple first order relations
\be\label{defdvpl}
\delta_\ipl = \Bary(\MyDelta_\ib,\MyDelta_\ir),\quad
v_\ipl=\Bary(v_\ib,v_\ir)\,,
\ee
where we defined \mbox{$\Bary(X_\ib,X_\ir)\equiv (RX_\ib+X_\ir)/(1+R)$}.
The conservation equations are then rewritten in a compact form as
\be\label{EqconservationClever}
{\rm C}_{\is}=\left( \frac{\MyDelta_{\is}}{1+w_\ipl}\right)'-k v_{\is}-3\Psi'=0\,.
\ee
In order to fully characterize this two fluids system we also need to
describe the differences between them by the entropy perturbation
\mbox{$S\equiv 3[(\rho_\ib/\bar \rho_\ib)^{1/3}- (\rho_\ir/\bar
\rho_\ir)^{1/4}]$}, and the velocity slip \mbox{$V^\mu \equiv
u^\mu_\ib-u^\mu_\ir$}. Both quantities vanish for adiabatic initial
conditions at any order in perturbations. We obtain at linear order
\be
(1+w_\ipl)S=\MyDelta_\ib-\MyDelta_\ir\,,\qquad V=v_\ib-v_\ir\,.
\ee
In the rest of this paper, we abbreviate $w_\ipl$ to $w$ and
$c_\ipl^2$ to $c^2$. The photons-baryons system is either described
with the intrinsic fluid variables
\mbox{$(\delta_\ir,\delta_\ib,v_\ib,v_\ir)$} or with the total fluid
variables \mbox{$(\delta_\ipl,v_\ipl,S,V)$} but the latter choice of variables is better suited for the TCA.\\

{\it Tight-coupling expansion: } From the definition~(\ref{defdvpl}), considering  $\Bary({\rm C}_\ib,{\rm
  C}_\ir)$ and $\Bary({\rm E}_\ib, {\rm E}_\ir)$ with
Eqs.~(\ref{EqconservationClever}) and (\ref{EqCollision1}) leads to the plasma equations
\bea
{\rm C}_\ipl&=&\HH \frac{R S}{(1+R)^2}\,,\\\label{EqsPlasma1}
{\rm E}_\ipl&=&k\left[\frac{RS}{3(1+R)^2}+\frac{1}{(1+R)}\frac{2\Theta_2}{5}\right].
\eea
Taking the differences ${\rm E}_\ib-{\rm E}_\ir$ and ${\rm C}_\ib
-{\rm C}_\ir$, we also obtain the entropy and velocity slip equations
\bea
&&\hspace{-0.3cm}S'-k V=0\,,\\\label{EqEntropy1}
&&\hspace{-0.3cm}-(1+R)/R\tau' V=\label{EqTCV}\\
&&\hspace{-0.3cm}\left[V'+\HH (v_\ipl+3 c^2
  V)+k\left(\frac{2}{5}\Theta_2-\frac{\delta_\ipl}{3(1+w)}+R
c^2S\right)\right]\nonumber.
\eea
The dynamics of the quadrupole $\Theta_2$ is inferred from the
Boltzmann hierarchy~\cite{Pitrou2008}. Combined with the dynamics of
$E_2$ we get
\be\label{Eqtheta2}
\tau'\Theta_2= 2 E_2'-\frac{4 \Theta_2'}{3}+k\left(-\frac{8 v_\ir}{9}-\frac{4
    \Theta_3}{7}+\frac{10
    E_3}{21}\right)
\ee
where we must then use that \mbox{$v_\ir=v_\ipl-RV/(1+R)$}. The same method
leads to 
\be\label{ConsE2}
\tau'E_2= -3E_2'+\frac{\Theta_2'}{3}+k\left(\frac{2 v_\ir}{9}+\frac{\Theta_3}{7}+\frac{5
    E_3}{7}\right)\,.
\ee
For higher order moments, the Boltzmann hierarchy leads directly to 
\bea
\tau' E_\ell&=&-E_\ell' +k\left[\frac{(\ell-1)(\ell+3)}{(\ell+1)(2 \ell+3)}E_{\ell+1}-\frac{\ell}{2
  \ell-1}E_{\ell-1}\right]\nonumber\label{Eqlastconstraint}\\
\tau' \Theta_\ell&=&-\Theta_\ell' +k\left(\frac{\ell+1}{2 \ell+3}\Theta_{\ell+1}-\frac{\ell}{2
  \ell-1}\Theta_{\ell-1}\right).
\eea
When combined with the Einstein equations to determine the
perturbations of the metric, these equations completely determine the
system. However, at early times $\tau' \propto a^{-2}$ and the restoring
forces can be huge and require very small numerical steps. The TCA
consists in not solving the dynamical equations for \mbox{($V$,$\Theta_{\ell\ge 2}$,$E_{\ell\ge 2}$)}  but to use instead the expressions~(\ref{EqTCV}-\ref{Eqlastconstraint}) multiplied by
$1/\tau'$  to obtain expressions for these variables as constraints which are functions of plasma
variables but also functions of themselves. To obtain a closed system out of
this infinite recursion, all variables are expanded in powers of
the TCA parameter $\epsilon$ in the form
\be\label{DefTCA}
X=\sum_{p=0}^\infty \epsilon^p X^{(p)}\qquad{\rm with}\qquad \epsilon\equiv \HH/\tau'\,,
\ee
and we use that $V^{(0)}=0$ and \mbox{$\Theta_{\ell \ge
  2}^{(n)}=E_{\ell\ge 2}^{(n)}=0$} for $n \le (\ell-2)$. The expansion~(\ref{DefTCA}) can be obtained by
replacing recursively the constraints in themselves, and getting rid
of the time derivative by using the plasma and entropy equations. We stop at the desired TCA order, gaining one order at each recursion, and then the constraints are
replaced in the plasma and entropy equations. We obtain finally a
\emph{closed} and \emph{first order} differential system up to the
desired TCA order. More precisely, the result is stated as\\
(*): Defining the dynamical and constrained variables \mbox{$\vec{{
      Y}}\equiv(v_\ipl,\delta_{\ipl},S,\Phi,\Psi)$} and \mbox{$\vec{ Z}\equiv(V,\Theta_2,E_2,\Theta_3,E_3,\dots)$}, then to solve for
their dynamics there exists two sets of matrices
${\bf M}_p$ and ${\bf N}_p$, the coefficients of which are sums of products of the $(k/\HH)^q$, the
$\HH^{-(q+1)}\dd^q H/\dd \eta^q$, and some functions of $R$ (this type
of matrix is called \emph{generic} in the following), such that the equations to solve are
\be\label{EqGenrecursion1}
\vec{ Y}'=\HH \sum_{p=0}^\infty \epsilon^p {\bf M}_p.\vec{ Y},\quad \vec{ Z}=\sum_{p=0}^\infty
\epsilon^p {\bf N}_p.\vec{ Y}
\ee
with ${\bf N}_0=0$. This implies in particular that there exists
generic matrices  ${\bf K}_p$ such that \mbox{$\vec{ Z}'=\HH \sum_{p=1}^\infty \epsilon^p {\bf
  K}_p.\vec{ Y}$} [obtained from \mbox{${\bf K}_p\equiv
(\epsilon^p {\bf
  N}_p)'/(\HH \epsilon^p)+\sum_{r=1}^{p}{\bf N}_r. {\bf M}_{p-r}$}].  If we define the variables \emph{up to  TCA-$n$} by \mbox{$X^{<n}\equiv \sum_{p=0}^n
\epsilon^p X^{(p)}$}, then  $\vec{ Y}^{<n}$ and $\vec{ Z}^{<n}$ are obtained from a truncation of the expansion~(\ref{EqGenrecursion1}).

\begin{figure}[!htb]
\includegraphics[width=8cm]{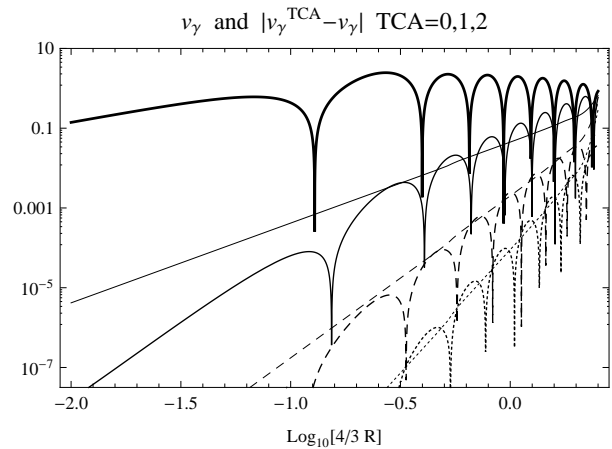}
\caption{Top thick line: exact $v_\ir$ with $k=0.2 {\rm
    Mpc}^{-1}$. The differences with the solutions up to TCA-$0,1,2$ are in continuous, dashed and dotted lines. The orders expected $(k/\tau')^{1,2,3}$, are depicted in thin lines of the same type.}\label{TC012}
\end{figure}

In order to show (*) we will determine these matrices
recursively. Hereafter ${\bf P}$, ${\bf Q}$, ${\bf U}$, ${\bf W}$,
${\bf B}$, and ${\bf C}$ are generic matrices. First, the dynamics of $\Psi$ from Einstein equations is of the form
$\Psi' =\HH {\bf P}.\vec{Y}$, but $\Phi$ is found from a constraint $\Phi-\Psi\propto
\Theta_2$ and its dynamics depends on $\Theta_2'$~\cite{Ma:1995ey}. Using
this, the plasma-entropy equations are of the form \mbox{$\vec{Y}'=\HH [{\bf Q}.\vec{ Y}+{\bf U}.\vec{ Z}+{\bf
  W}.\vec{Z}'/\HH]$}, the dependence in $\vec{Z}'$ being just on ${\Theta^\ir_2}'$. The ${\bf M}_p$ are not independent since ${\bf M}_0={\bf
  Q}$ and ${\bf M}_{p}={\bf U}.{\bf N}_{p}+{\bf W}.{\bf K}_{p}$ if
$p>0$.  Hence, the ${\bf K}_p$ and ${\bf M}_p$ are deduced from the ${\bf N}_p$. For $p=0$, \mbox{${\bf M}_0={\bf Q}$} and
\mbox{${\bf N}_0={\bf K}_0 =0$}. At TCA-0, it is as if there was a single perfect fluid.
Now, if the matrices are known up to a given $p\ge 0$, that is if we know
$\vec{Y}^{'<p}$ and $\vec{Z}^{<p}$ (and thus $\vec{Z}^{'<p}$), then we can find ${\bf K}_{p+1}$, and using that the constraints for $\vec{
  Z}$ is of the form
\be\label{EqStructureConstraint}
\vec{ Z}=\epsilon\left[{\bf A}.\vec{ Z}'/\HH+{\bf B}.\vec{ Z}+{\bf
    C}.\vec{ Y}\right]
\ee
where ${\bf A}$ is a constant matrix, we replace $\vec{ Y}^{'<p}$, $\vec{ Z}^{<p}$ and  $\vec{
  Z}^{'<p}$ in the right hand side. We then get ${\bf
  N}_{p+1}$ from
\be
{\bf N}_{p+1}={\bf  A}.{\bf K}_p+{\bf B}.{\bf N}_p
\ee
for $p>1$ and ${\bf N}_1={\bf C}$ otherwise, from which we find ${\bf M}_{p+1}$. At early times
$(\tau')'=-2\HH\tau'$, and $R'=\HH R$, and this is why matrices are
of generic type. For a realistic case, neutrinos an cold dark matter
should be considered and added to $\vec{Y}$. Note also that when recombination starts, $\tau'$
is no more scaling like $\propto a^{-2}$ and one must use its complete
expression. However, $(\tau')'$ factors appear only at TCA-2. The public code CAMB~\cite{CAMB} uses
the TCA-1  with intrinsic fluid variables instead of
total fluid variables, and such factors appear already at that
order. With this algorithm, we can find the equations up to a given TCA-$n$ if one uses abstract calculus to perform the recursion. Our approach clarifies the
method adopted in Ref.~\cite{Doran:2005ep} where (*) is implicitely used up to TCA-2. 
At early times, it can be used to avoid instablities due to the high
interaction rate. Using higher orders of the TCA enables to improve
the accuracy, given that up to TCA-$n$ th precision is roughly of order $(k/\HH
\epsilon)^n=(k/\tau')^n$ for large modes and $\epsilon^n$ otherwise. In Fig.~\ref{TC012} we plot
the error between the full numerical integration and the successive orders of the TCA for $v_\ir$.

{\it Baryon acoustic oscillations:} Qualitative features of the BAO can be inferred by computing the
evolution of $\delta_\ir$ at TCA-1. Considering the combination ${\rm
  C}_\ir'+k{\rm E}_\ir$ and
expressing $\Theta_2$ up to TCA-1 and $V$ up to TCA-2 in order to obtain
$V/\epsilon$ up to TCA-1, and further ignoring the variations of the
potentials, we obtain
\bea
&&\hspace{-0.38cm}\delta_\ir''\left[1-\frac{\epsilon  R^2}{(1+R)^2}\right]
+k^2 c^2
\delta_\ir\left[1+\frac{\epsilon R^2}{1+R}\left(\frac{2+R}{1+R}-r\right) \right]\nonumber\\
&&\hspace{-0.38cm}+\delta_\ir'\left[\frac{\HH  R}{1+R}+\epsilon \frac{k^2}{\HH} c^2
  \left(\frac{16}{15}+\frac{R^2}{1+R}\right)+\dots\right]=-\frac{4}{3}k^2
\Phi\,,\nonumber
\eea
where $r\equiv (\ln \tau')'/\HH$ and  the dots in the friction term represent terms which are suppressed by a factor
$(\HH/k)^2$ with respect to the dominant TCA-1 term. We thus recover
that up to TCA-1, the viscosity is damping the oscillations~\cite{Hu:1995en}, but we also
find that the pseudo-period of oscillations, and thus the sound
horizon, is slightly modified by the TCA-1 terms. However for large modes, $k/\tau'>1$ before $\epsilon>1$ and the effect
of viscosity is expected to be dominant over the modified pseudo-period variation. This approach is of limited precision, since
the TCA cannot be trusted around recombination, and only the full numerical integration makes sense.

\subsection{Second order cosmological perturbations: vorticity}

At second order, the same procedure can be followed. Given the precision
that we need to reach, the TCA-0 is sufficient, and the equations are
those of a perfect fluid. They can be found e.g. in
Ref.~\cite{Pitrou:2008ak} (with a different definition of the velocity
perturbation). There is however a case where the TCA-0 is not
sufficient. Indeed, vorticity is not generated for a perfect fluid~\cite{Lu:2008ju} and
thus not at TCA-0~\cite{Kobayashi:2007wd}. One needs then to consider
the equations up to TCA-1 at least in order to describe the generation of
vorticity. The Euler equation for a perfect fluid reads
\bea\label{Eulerappendix}
{\rm E}^\is_\iT &\equiv&  {v^{\is}_\iT}'+(1-3 {c_\is^2})\HH v^{\is}_\iT+\frac{c_\is^2}{1+w_\is}\partial_\iB \delta_\is+\partial_\iB \Phi -4 \Psi' v^\is_\iT\nonumber\\
&+&\frac{1+c_\is^2}{1+w_\is}\left[ (\delta_\is v^\is_\iT)'+\HH(1-3 w_\is)\delta_\is v^\is_\iT+\delta_\is\partial_\iB \Phi\right]\nonumber\\
&+&\partial_\jB(v^\is_\iT v_\is^\jT)-(\Phi+\Psi)\left[{v^\is_\iT}'+\HH(1-3 c_\is^2)v^\is_\iT\right]-\partial_\iB(\Phi^2)\nonumber\\
&+&\Psi\left[{v^{\is}_\iT}'+(1-3c_\is^2)\HH v^{\is}_\iT+\frac{c_\is^2}{1+w_\is}\partial_\iB \delta_\is+\partial_\iB \Phi \right]\nonumber\\
&+&\frac{{(c_\is^2)}'}{1+w_\is}\delta_\is u^\is_\iT-\frac{{(c_\is^2)}'}{6\HH(1+w_\is)^2}\partial_\iB(\delta_\is^2)={\cal C}^{v_\is}_{\iT}\,.
\eea
The vorticity is defined by \mbox{$\omega^\is_{\mu\nu} \equiv
{\perp^{\is}}_\mu^{\,\,\alpha} {\perp^{\is}}_\nu^{\,\,\beta}
\nabla_{[\alpha }u^{\is}_{\beta]}$} where
\mbox{$X_{[\mu\nu]}\equiv\frac{1}{2}(X_{\mu\nu}-X_{\nu\mu})$}, and the
projector is given by ${\perp^\is}_\mu^{\,\,\nu}\equiv
\delta_\mu^\nu+u^\is_\mu u_\is^\nu$. There is no vorticity a first order since we have discarded the vector modes, but at second order its
expression is
\be\label{EqExpressionomega}
a \omega^\is_{\iT \jT} = \partial_{[\iB} v^\is_{\jT]}+  v^\is_{[\iT} \partial_{\jB]} (\Psi+\Phi) +v^\is_{[\iT} {v^\is_{\jT]}}'\,.
\ee
For a collisionless perfect fluid, that is if ${\cal C}^{v_\is}_{\iT}=0$, then from the Euler equation, we can show that the evolution of vorticity is dictated by
\be
{\rm \Omega}^\is_{\iT \jT}\equiv{\omega^\is_{\iT
    \jT}}'+(2-3c_\is^2)\HH \omega^\is_{\iT \jT}=0
\ee
which implies that $[\bar \rho_\is(1+w_\is)a^5 \omega^\is_{\iT \jT}]$ is conserved. At second order, the velocity of the plasma is given by
\be
\hspace{-0.0cm}\Bary(1+\delta_\ib,1+\delta_\ir)u^\mu_\ipl\equiv\Bary[(1+\delta_\ib)u^\mu_\ib,(1+\delta_\ir)u^\mu_\ir].
\ee
Using this, we find as expected that at TCA-0 the baryons-photons plasma behaves like a
perfect fluid, since $\Bary({\rm E}^{\ib(0)}_\iT,{\rm E}^{\ir(0)}_\iT)={\rm  E}^\ipl_\iT$, and the plasma vorticity satisfies ${\rm \Omega}^{\ipl(0)}_{\iT\jT}=0$. However in full generality
\be\label{EqMalik}
{\rm \Omega}^\ipl_{\iT\jT}+\Omega^\Pi_{\iB\jB}+\frac{R}{(1+R)^2}\left[\frac{\partial_{[\iT}\delta_\ipl^{\rm com}\partial_{\jT]}S}{3(1+w_\ipl)}+\partial_{[\iT}(\partial_\kB V^\kT) \partial_{\jT]}V \right]=0\nonumber 
\ee
with $\delta_\ipl^{\rm com}\equiv \delta_\ipl-3 \HH(1+w_\ipl) v_\ipl$,
and where $\Omega^\Pi_{\iB\jB}=0$ if for photons we neglect the
anisotropic stress and use a perfect fluid description. This equation without the last term (which is related to the quadrupole generated by
the mixing of the fluids) matches the expression of Ref.~\cite{Christopherson:2009bt}, given
that the non-adiabatic pressure perturbation is 
\be
\delta P_{\rm nad}\equiv c_\ir^2 \bar \rho_\ir \delta_\ir-c_\ipl^2
\bar \rho_\ipl \delta_\ipl =-\bar \rho_\ipl\frac{R (1+w_\ipl)}{3(1+R)^2}S\,.
\ee
However we do need to consider the anisotropic stress at TCA-$(n\ge 1)$. For
completeness, its contribution is
\bea
&&\hspace{-0.4cm}\Omega^\Pi_{\iB\jB}\equiv \frac{3 c^2}{4}\left[-
\frac{1+c^2}{1+w}\partial_{[\iT}\delta_\ipl \partial^{\kT}\Pi^\ir_{\jT]\kT}+\frac{v^\ipl_{[\iT}(c^2
 \partial^{\kT}\Pi^\ir_{\jT]\kT})^{'}}{c^2}+\right.\nonumber\\
&&\hspace{-0.5cm}\left.\frac{R \partial_{[\iT}S \partial^{\kT}\Pi^\ir_{\jT]\kT}}{3(1+R)^2} +\partial_{[\iT} \partial^\kT[(1+\delta_\ir) \Pi^\ir_{\jT]k}]+\partial_{[\iT}\partial^\kT (\Phi-3\Psi) \Pi^\ir_{\jT]k}\right]\nonumber
\eea

%
In this paper, we have formulated the TCA in the total fluid variables since it can then be extended easily up to any
order through a recursions on the equations. Higher orders
in the TCA can be used to speed up and refine linear Boltzmann codes~\cite{CAMB,Doran:2005ep}. At second order, a
TCA-0 solution is generally sufficient, e.g. for the
computation of non-Gaussianity generated by non-linear effects~\cite{2010JCAP...07..003P}. 
However, the vorticity in the baryons-photons fluid is generated at
least at TCA-1, not only from gradients of non-adiabatic pressure
perturbations (see Refs.~\cite{Kobayashi:2007wd,Christopherson:2009bt}), but also from gradients of
the anisotropic stress of photons. This becomes relevant for the
numerical estimation of the seed magnetic field created by vortical
currents~\cite{Fenu2011} since it is related to the existence of
vorticity.\\
{\it Acknowledgements:}  C. P. is supported by STFC (UK) grant ST/H002774/1 and thanks
J.-P. Uzan and R. Maartens for discussions on the topic. 

\bibliography{bibTC}

\begin{thebibliography}{16}
\expandafter\ifx\csname natexlab\endcsname\relax\def\natexlab#1{#1}\fi
\expandafter\ifx\csname bibnamefont\endcsname\relax
  \def\bibnamefont#1{#1}\fi
\expandafter\ifx\csname bibfnamefont\endcsname\relax
  \def\bibfnamefont#1{#1}\fi
\expandafter\ifx\csname citenamefont\endcsname\relax
  \def\citenamefont#1{#1}\fi
\expandafter\ifx\csname url\endcsname\relax
  \def\url#1{\texttt{#1}}\fi
\expandafter\ifx\csname urlprefix\endcsname\relax\def\urlprefix{URL }\fi
\providecommand{\bibinfo}[2]{#2}
\providecommand{\eprint}[2][]{\url{#2}}

\bibitem[{\citenamefont{de~Bernardis et~al.}(2000)}]{Bernardis:2000gy}
\bibinfo{author}{\bibfnamefont{P.}~\bibnamefont{de~Bernardis}}
  \bibnamefont{et~al.} (\bibinfo{collaboration}{Boomerang}),
  \bibinfo{journal}{Nature} \textbf{\bibinfo{volume}{404}},
  \bibinfo{pages}{955} (\bibinfo{year}{2000}), \eprint{astro-ph/0004404}.

\bibitem[{\citenamefont{Komatsu et~al.}(2010)}]{Komatsu:2010fb}
\bibinfo{author}{\bibfnamefont{E.}~\bibnamefont{Komatsu}} \bibnamefont{et~al.}
  (\bibinfo{year}{2010}), \eprint{1001.4538}.

\bibitem[{\citenamefont{Eisenstein et~al.}(2005)}]{Eisenstein:2005su}
\bibinfo{author}{\bibfnamefont{D.~J.} \bibnamefont{Eisenstein}}
  \bibnamefont{et~al.} (\bibinfo{collaboration}{SDSS}),
  \bibinfo{journal}{Astrophys. J.} \textbf{\bibinfo{volume}{633}},
  \bibinfo{pages}{560} (\bibinfo{year}{2005}), \eprint{astro-ph/0501171}.

\bibitem[{\citenamefont{Peebles and Yu}(1970)}]{Peebles1970}
\bibinfo{author}{\bibfnamefont{P.~J.~E.} \bibnamefont{Peebles}}
  \bibnamefont{and} \bibinfo{author}{\bibfnamefont{J.~T.} \bibnamefont{Yu}},
  \bibinfo{journal}{Astrophys. J.} \textbf{\bibinfo{volume}{162}},
  \bibinfo{pages}{815} (\bibinfo{year}{1970}).

\bibitem[{\citenamefont{Ma and Bertschinger}(1995)}]{Ma:1995ey}
\bibinfo{author}{\bibfnamefont{C.-P.} \bibnamefont{Ma}} \bibnamefont{and}
  \bibinfo{author}{\bibfnamefont{E.}~\bibnamefont{Bertschinger}},
  \bibinfo{journal}{Astrophys. J.} \textbf{\bibinfo{volume}{455}},
  \bibinfo{pages}{7} (\bibinfo{year}{1995}), \eprint{astro-ph/9506072}.

\bibitem[{\citenamefont{Kobayashi et~al.}(2007)\citenamefont{Kobayashi,
  Maartens, Shiromizu, and Takahashi}}]{Kobayashi:2007wd}
\bibinfo{author}{\bibfnamefont{T.}~\bibnamefont{Kobayashi}},
  \bibinfo{author}{\bibfnamefont{R.}~\bibnamefont{Maartens}},
  \bibinfo{author}{\bibfnamefont{T.}~\bibnamefont{Shiromizu}},
  \bibnamefont{and}
  \bibinfo{author}{\bibfnamefont{K.}~\bibnamefont{Takahashi}},
  \bibinfo{journal}{Phys. Rev.} \textbf{\bibinfo{volume}{D75}},
  \bibinfo{pages}{103501} (\bibinfo{year}{2007}), \eprint{astro-ph/0701596}.

\bibitem[{\citenamefont{Maeda et~al.}(2009)\citenamefont{Maeda, Kitagawa,
  Kobayashi, and Shiromizu}}]{Maeda2008}
\bibinfo{author}{\bibfnamefont{S.}~\bibnamefont{Maeda}},
  \bibinfo{author}{\bibfnamefont{S.}~\bibnamefont{Kitagawa}},
  \bibinfo{author}{\bibfnamefont{T.}~\bibnamefont{Kobayashi}},
  \bibnamefont{and}
  \bibinfo{author}{\bibfnamefont{T.}~\bibnamefont{Shiromizu}},
  \bibinfo{journal}{Class. Quant. Grav.} \textbf{\bibinfo{volume}{26}},
  \bibinfo{pages}{135014} (\bibinfo{year}{2009}), \eprint{0805.0169}.

\bibitem[{\citenamefont{Pitrou}(2009)}]{Pitrou2008}
\bibinfo{author}{\bibfnamefont{C.}~\bibnamefont{Pitrou}},
  \bibinfo{journal}{Class. Quant. Grav.} \textbf{\bibinfo{volume}{26}},
  \bibinfo{pages}{065006} (\bibinfo{year}{2009}), \eprint{0809.3036}.

\bibitem[{\citenamefont{Lewis and Challinor}()}]{CAMB}
\bibinfo{author}{\bibfnamefont{A.}~\bibnamefont{Lewis}} \bibnamefont{and}
  \bibinfo{author}{\bibfnamefont{A.}~\bibnamefont{Challinor}},
  \emph{\bibinfo{title}{Camb}}, \urlprefix\url{http://camb.info}.

\bibitem[{\citenamefont{Doran}(2005)}]{Doran:2005ep}
\bibinfo{author}{\bibfnamefont{M.}~\bibnamefont{Doran}},
  \bibinfo{journal}{JCAP} \textbf{\bibinfo{volume}{0506}}, \bibinfo{pages}{011}
  (\bibinfo{year}{2005}), \eprint{astro-ph/0503277}.

\bibitem[{\citenamefont{Hu and Sugiyama}(1996)}]{Hu:1995en}
\bibinfo{author}{\bibfnamefont{W.}~\bibnamefont{Hu}} \bibnamefont{and}
  \bibinfo{author}{\bibfnamefont{N.}~\bibnamefont{Sugiyama}},
  \bibinfo{journal}{Astrophys. J.} \textbf{\bibinfo{volume}{471}},
  \bibinfo{pages}{542} (\bibinfo{year}{1996}), \eprint{astro-ph/9510117}.

\bibitem[{\citenamefont{Pitrou et~al.}(2008)\citenamefont{Pitrou, Uzan, and
  Bernardeau}}]{Pitrou:2008ak}
\bibinfo{author}{\bibfnamefont{C.}~\bibnamefont{Pitrou}},
  \bibinfo{author}{\bibfnamefont{J.-P.} \bibnamefont{Uzan}}, \bibnamefont{and}
  \bibinfo{author}{\bibfnamefont{F.}~\bibnamefont{Bernardeau}},
  \bibinfo{journal}{Phys. Rev.} \textbf{\bibinfo{volume}{D78}},
  \bibinfo{pages}{063526} (\bibinfo{year}{2008}), \eprint{0807.0341}.

\bibitem[{\citenamefont{Lu et~al.}(2009)\citenamefont{Lu, Ananda, Clarkson, and
  Maartens}}]{Lu:2008ju}
\bibinfo{author}{\bibfnamefont{T.~H.-C.} \bibnamefont{Lu}},
  \bibinfo{author}{\bibfnamefont{K.}~\bibnamefont{Ananda}},
  \bibinfo{author}{\bibfnamefont{C.}~\bibnamefont{Clarkson}}, \bibnamefont{and}
  \bibinfo{author}{\bibfnamefont{R.}~\bibnamefont{Maartens}},
  \bibinfo{journal}{JCAP} \textbf{\bibinfo{volume}{0902}}, \bibinfo{pages}{023}
  (\bibinfo{year}{2009}), \eprint{0812.1349}.

\bibitem[{\citenamefont{Christopherson
  et~al.}(2009)\citenamefont{Christopherson, Malik, and
  Matravers}}]{Christopherson:2009bt}
\bibinfo{author}{\bibfnamefont{A.~J.} \bibnamefont{Christopherson}},
  \bibinfo{author}{\bibfnamefont{K.~A.} \bibnamefont{Malik}}, \bibnamefont{and}
  \bibinfo{author}{\bibfnamefont{D.~R.} \bibnamefont{Matravers}},
  \bibinfo{journal}{Phys. Rev.} \textbf{\bibinfo{volume}{D79}},
  \bibinfo{pages}{123523} (\bibinfo{year}{2009}), \eprint{0904.0940}.

\bibitem[{\citenamefont{{Pitrou} et~al.}(2010)\citenamefont{{Pitrou}, {Uzan},
  and {Bernardeau}}}]{2010JCAP...07..003P}
\bibinfo{author}{\bibfnamefont{C.}~\bibnamefont{{Pitrou}}},
  \bibinfo{author}{\bibfnamefont{J.}~\bibnamefont{{Uzan}}}, \bibnamefont{and}
  \bibinfo{author}{\bibfnamefont{F.}~\bibnamefont{{Bernardeau}}},
  \bibinfo{journal}{JCAP} \textbf{\bibinfo{volume}{1007}}, \bibinfo{pages}{003}
  (\bibinfo{year}{2010}), \eprint{1003.0481}.

\bibitem[{\citenamefont{Fenu et~al.}(in preparation)\citenamefont{Fenu, Pitrou,
  and Maartens}}]{Fenu2011}
\bibinfo{author}{\bibfnamefont{E.}~\bibnamefont{Fenu}},
  \bibinfo{author}{\bibfnamefont{C.}~\bibnamefont{Pitrou}}, \bibnamefont{and}
  \bibinfo{author}{\bibfnamefont{R.}~\bibnamefont{Maartens}} (\bibinfo{year}{in
  preparation}).

\end{thebibliography}

\end{document}